# PROBLEM BASED LEARNING AND ITS IMPLEMENTATIONS

SEIBU MARY JACOB[†]

School of Business, Swinburne University of Technology (Sarawak Campus),
Kuching, Sarawak, Malaysia
sjacob@swinburne.edu.my

BIJU ISSAC

School of IT & Multimedia, Swinburne University of Technology (Sarawak Campus),
Kuching, Sarawak, Malaysia
bissac@swinburne.edu.my

In an era where learning is considered a problem, we decided to go for problems for the sake of learning! The purpose of this study was to throw light on the issues involved in two forms of PBL viz., Case Study Based PBL and Research Based PBL The influence of course and subject on the learning method was investigated. Students' perceptions and concerns were analyzed using questionnaires. A slight variation was found among the different courses of Engineering, Business and IT, as also among the sample subjects considered. It is concluded that careful and systematic introduction to the case study based learning can be progressively led to the Research based learning as the student progresses from the first semester to the final semester of a graduate degree course or from graduate degree course to post graduate degree course.

## 1. Introduction

In their desperate quest to make curriculum and learning more relevant to their students, educators explore on new delivery systems. Without doubt, the possibilities are great and the language lofty when addressing the latest wave of 'Problem-based Learning'. PBL outcries for authentic learning, real world problems, constructivist classrooms, performance assessment. Here we are investigating into the implementation of PBL in the form of the Case Study Based Learning and Research Based Learning in our University, and echoing the voices of our University students. We are trying to explore the statement 'There's nothing so practical as good theory, and nothing so theoretically interesting as good practice[1].' We echo the opinions of other educators regarding the fact that, PBL exists in the eye of the beholder[2].So we believe it's important to write about PBL including sufficient details to enable us to see how we compare with others.

---

[†] Swinburne University of Technology (Sarawak Campus), Level 1, State Complex, Jalan Simpang Tiga, 93576 Kuching, Sarawak, Malaysia.

## 2. PBL & Authentic Learning

PBL starts off with the introduction of an ill-structured problem which becomes the focus of learning. If PBL has to ensure authentic learning, it should be based on the Experiential Learning Cycle (ELC)[5]. This can be categorized on three levels of increasing authenticity & complexity: Academic challenges, which may be student work, structured as a problem arising directly from an area of the subject, to promote greater understanding of the area, Scenario challenges, through which students begin to see themselves in real life roles as they develop knowledge and skills for success in University and beyond and Real life problems, through which students move outside the classroom, take action on issues, and have a tangible impact on their communities. These are real problems in need of real solutions by real people or organizations. ELC speaks of both student and teacher dimensions. The student roles are Engagement (of the students in the problem carefully crafted by the teacher), Exhibition (of the concrete product developed as the evidence of learning) and Examination (of the work and reflection on the concepts learned). The teacher roles are of a Designer (to carefully design the problem and lead the students into the engagement phase. Here, the standards of quality are pre-established), Coach (to develop skills, shape strategies, find appropriate resources and letting students own their successes and failures) and Feedback giver (to create structures within which students stand to reflect and assess their products, processes, level of understanding based on the pre-established standards of quality).

## 3. Research Based PBL

Under Research based PBL the details of two undergraduate students who did a research project for around five months under the supervision of one of the authors is presented. This research project had two profound results. Firstly, their research paper with the all the results that they found was accepted in IEEE International Conference on Networks (ICON 2005), to be held in Kuala Lumpur during November 2005. Secondly, one of the students is now doing his post graduate degree in NTU, Singapore whose topic of research is an extension of what he did under the research based project learning earlier. The problems given were to practically setup a test bed and analyze the security vulnerabilities of an IEEE802.11b wireless network. The students were successful in setting up the test bed with encryption and RADIUS authentication and analyzing it.

### 3.1. Cycle of Steps involved

The cycle of steps involved in Research based PBL can be outlined as follows: Step1: Problem and expected output defined by the supervisor → Step 2: General literature review to learn the general problem scenario → Step 3: Divide the large problem domain

Problem Based Learning and its Implementationsinto small sections/domains → Step 4: More specific literature review on specific domains (80%) or minimal discussion with the supervisor (20%) → Step 5: Design of the smaller domains → If failure, go back to step 4 → Step 6: Implementation of the smaller domains → If failure, go back to step 4 → Step 7: Integration of the smaller domains to get the full problem domain → If failure, go back to step 4 → Step 8: Test and trouble shoot → If failure, go back to step 4 → Step 9: Document the results.

**3.2. Qualitative Analysis of feedbacks**

The students were given a questionnaire which had both open and closed questions. Some of the positive and negative answers for the open questions given are as follows: Positive feedbacks – It gives more freedom to explore and find out new knowledge. In some cases they can actually discover and understand more on the related topics which normally they can't get in their lecturers notes. It allows them to get a better understanding of the topic through literature review and practical implementation. Negative feedbacks – Some times, there is too much doubt when students learn by research method due to too much information that students can obtain from the internet or other places. The whole process can be very time consuming and can test their patience.

**3.3 Quantitative Analysis of Choices made**

Nearly nine closed questions were given with multiple choices, and generally both the students were quite positive about the research based PBL. They preferred a mix of classroom learning and Research Based learning in a subject's curriculum. The reasons chosen for favoring the Research Based learning mode were: Free from the rigid classroom atmosphere, intellectually challenging, can refer to any relevant resource. The advantages they had picked up were: Learning happens in steps and not just for final exams, instills confidence and gave a sense of empowerment. The teacher's roles were picked up as: The teacher is the source which initiates the process, Teacher intervenes (& clarifies doubts) with minimal help in between the stages.

**4. Case Study Based PBL**

Some case study examples from degree subjects are presented below, with their initial knowledge of the problem being thin. Our student samples varied from first to fourth year degree students in Engineering, IT and Business.

**4.1. Case Studies of Sample subjects**

4.1.1. Software Development 2

Software Development 2 (SD2) is a second level Java programming paper which the second year engineering students undertake after finishing the pre-requisite of Software Development 1. Two assignments were given as follows: Firstly an Assessment

551

Problem Based Learning and its Implementations

Processing Program had to be written in Java to read from two input files and they need to process these read values from two input files and write them into an output assessment file in a formatted fashion. Secondly, a Hangman Game in Java had to be written in Java through Graphical User Interface using Java Swing and Event handling options. This is a guess word game where the player could win or lose and if he loses, a hangman picture is drawn.

4.1.2. Introduction to Software Engineering

In Introduction to Software Engineering (ISE), the third year engineering students had to do two assignments as follows: Firstly, to create the Software Requirements Specification (SRS) document from a template given for a software development project that a fictitious company gets. Secondly, to create the Software Test Document (STD) from a template for the SRS requirements done above so that testing of the software is done in a comprehensive way.

4.1.3. C++ for Programmers

In C++ for Programmers, the final year engineering students were given one assignment as given: To create a menu based Invoice and Billing program in C++ for a Stall that sells ice creams and milk products, so that they can come out with the invoice and the final bill for the customers who buy different products, including the sales tax calculation.

4.1.4. Quantitative Analysis in Business

A compulsory subject for the first year degree business students, Quantitative Analysis (Q.A.) involves a case study, where the students are provided with a data set from a real survey. The first part is to summarize the data set using descriptive statistics and the second part, to describe using inferential statistics. The expected output is a full fledged Business Report, with all the formalities of Executive Summary, Introduction, Analysis and Conclusion.

4.2  Students' Perceptions-Qualitative Analysis of feedbacks

A survey was conducted among the students who have taken one (or more) among the above listed papers, using a tested questionnaire consisting of open ended and closed ended questions. Below is summarized the open ended comments: Positive feedbacks – It helped them with understanding than memorizing materials. "Putting it into practice is what helps to remember the knowledge acquired", says a student .They acknowledged it was good for their future and made them more independent and responsible. It made them think and encouraged independent learning. Negative feedbacks – It meant more work for students, as group case study always has 'sleeping' partners. It made their heads spin and was quite demanding on their regular study time. Some times it took too much time to find out what case study requires. Lazy students might neglect their work and depend on others.

**4.3 Students' Perceptions-Quantitative Analysis of feedbacks**

The students' feedbacks showed that close to a majority of the Q.A. and ISE students have viewed the CSBL as okay, followed by a lower majority as pretty difficult. In C++, almost equal percentages of students have viewed as very much needed and pretty difficult. A slightly higher percentage of SD2 students have viewed CSBL as pretty difficult compared to a lesser percentage as very much needed. The closed ended questions in the survey concerned the students' view of case study based learning, its advantages and the teacher's role. The reliability of this part, questioning whether the aspects were perceived by the students as intended by the designers of the assessment, was moderately high (coefficient alpha = 0.67). A statistical analysis of the perceptions was done using SPSS. The difference among the Business, Engineering & IT streams, was statistically significant regarding the rating on case study based learning (Cramer's V=0.210, p<0.05). A high percentage of engineering students rated it as excellent compared to the others as in figure 1. The different aspects analyzed were as follows: View on Case Study based Learning – Across the different streams of Business, IT and Engineering, there was not much variation in views. A high proportion in each stream viewed it as 'okay' in preference to 'very much needed'. A slightly low proportion viewed it as 'pretty difficult'. Choice of Learning – The students welcome a mix of classroom based learning in preference to an exhaustive classroom teaching based curriculum or Case study based curriculum. Reasons & advantages in Case Study based Learning – Among the advantages noted were the possibilities of referring to any relevant source, gaining knowledge and scoring marks (on an average scale), helping one to learn more than what is required for exams. They also voiced the merits that learning happens in steps and not just for final exams and it encourages discussion with peers and team. *Teacher's role* – The teacher's role was viewed highly in 3 aspects: Teacher intervenes (and clarifies doubts) with minimal help in between the stages, Teacher encourages student to take up responsibility to learn, Teacher can keep the student informed of his progress in learning.

**5. Conclusions**

We were trying to put ourselves to test before John Dewey's quote: 'Thinking is doing and we must give the learners the challenge of doing something with their thinking[6]'. Our experience and student feed backs gave us the following conclusive information: PBL students use learning resources differently, they express more confidence in their information-seeking skills, they said they enjoyed the problem and wished they had more time to research more. Others felt the limited time kept them focused on one thing and did not allow time to get bored with the subject, they felt they were given the freedom to come up with solutions that made sense to them without the fear of being "wrong" and they became better at thinking about multiple solutions rather than jumping to their first conclusions. What can be encouraged from our experience is to start off with the case study based learning in the initial semesters of the undergraduate courses, to train and mature students in independent and self -directed learning. Then as they approach their final year, the research based learning can be introduced as a higher version of what they have been experiencing in case study based learning.

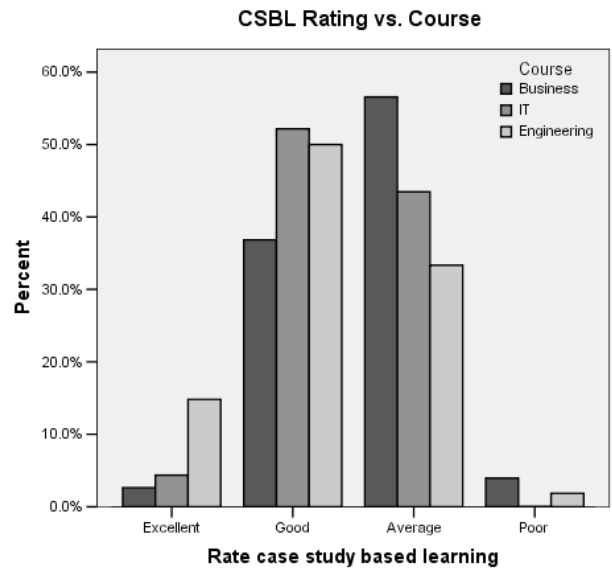

Fig.1. A high percentage of engineering students rated CSBL as excellent compared to the others. A general majority rates it as good.

## References

1. Gaffney, J.S., & Anderson, R.C. Two-tiered Scaffolding: Congruent processes of teaching and learning. In Problem Based Learning: An instructional model and its constructivist framework, eds. John R. Savery & Thomas M. Duffy (CLRT Technical Report No. 16-01, Indiana University, 2001).
2. Trudy W. Banta, Karen E. Black, Kimberly A. Kline, PBL 2000 Plenary Address Offers Evidence For and Against Problem Based Learning (PBL Insight,Vol.3 No.3, http://www.samford.edu./pbl)
3. Vernon D.T.A., & Blake, R.L., Does problem –based learning work? A Meta analysis of evaluative research, (Academic Medicine, 68(7), 1993), pp 550-563.
4. Barrows, H.S., A Taxonomy of Problem Based Learning Methods, (Medical Education, 20, 481-486)
5. Rick Gordon, Balancing Real–World Problems with Real–World Results,(Phi Delta Kappan)
6. James Neill, Summary of John Dewey's "Experience & Education" http://www.wilderdom.com/experiential/DeweyExperienceEducation.htm